\documentclass{article}

\PassOptionsToPackage{numbers, compress}{natbib}
\usepackage[preprint]{neurips_2023}

\usepackage[utf8]{inputenc} 
\usepackage[T1]{fontenc}    
\usepackage{hyperref}       
\usepackage{url}            
\usepackage{booktabs}       
\usepackage{amsfonts}       
\usepackage{nicefrac}       
\usepackage{microtype}      
\usepackage{xcolor}         
\usepackage{amsmath}
\usepackage{amssymb}
\usepackage{mathtools}
\usepackage{amsthm}
\usepackage{multirow}
\usepackage{wrapfig}
\usepackage[capitalize,noabbrev]{cleveref}
\usepackage{graphicx}
\usepackage{subfigure} 
\usepackage{float}
\usepackage{algorithm} 
\usepackage{algorithmic}
\usepackage{textcomp} 
\usepackage{listings} 
\usepackage{enumitem}
\usepackage{utfsym}
\usepackage{times}
\usepackage{latexsym}
\usepackage{fancyvrb}
\usepackage{tablefootnote}
\usepackage{CJKutf8}

\newcommand{\commentout}[1]{}

\title{FireRedChat: A Pluggable, Full-Duplex Voice Interaction System with Cascaded and Semi-Cascaded Implementations}

\author{
FireRed Team\thanks{Authors (alphabetical order): Junjie Chen, Yao Hu, Junjie Li, Kangyue Li, Kun Liu, Wenpeng Li, Xu Li, Ziyuan Li, Feiyu Shen, Xu Tang, Manzhen Wei, Yichen Wu, Fenglong Xie, Kaituo Xu, Kun Xie. Corresponding Author: Fenglong Xie (fenglongxie@xiaohongshu.com)} \\
Xiaohongshu \\
}

\begin{document}
\maketitle

\begin{abstract}

Full-duplex voice interaction allows users and agents to speak simultaneously with controllable barge-in, enabling lifelike assistants and customer service. Existing solutions are either end-to-end, difficult to design and hard to control, or modular pipelines governed by turn-taking controllers that ease upgrades and per-module optimization; however, prior modular frameworks depend on non-open components and external providers, limiting holistic optimization. In this work, we present a complete, practical full-duplex voice interaction system comprising a turn-taking controller, an interaction module, and a dialogue manager. The controller integrates streaming personalized VAD (pVAD) to suppress false barge-ins from noise and non-primary speakers, precisely timestamp primary-speaker segments, and explicitly enable primary-speaker barge-ins; a semantic end-of-turn detector improves stop decisions. It upgrades heterogeneous half-duplex pipelines, cascaded, semi-cascaded, and speech-to-speech, to full duplex. Using internal models, we implement cascaded and semi-cascaded variants; the semi-cascaded one captures emotional and paralinguistic cues, yields more coherent responses, lowers latency and error propagation, and improves robustness. A dialogue manager extends capabilities via tool invocation and context management. We also propose three system-level metrics, barge-in, end-of-turn detection accuracy, and end-to-end latency, to assess naturalness, control accuracy, and efficiency. Experiments show fewer false interruptions, more accurate semantic ends, and lower latency approaching industrial systems, enabling robust, natural, real-time full-duplex interaction. Demos: \url{https://fireredteam.github.io/demos/firered_chat}.

\end{abstract}

\section{Introduction}
Voice interaction refers to a form of human-computer interaction in which speech serves as the user’s input modality, the system recognizes and interprets the input, and then generates an appropriate spoken response. Voice interaction plays a critical role in AI assistants, automated customer service, real-time translation, and human-robot interaction. It is commonly categorized by communication mode into simplex, half-duplex (turn-based), and full-duplex~\cite{ma2025language}. In simplex interaction, only one party is able to speak; in half-duplex interaction, the user and the agent must take turns, waiting for the other to finish before speaking (i.e., turn-based dialogue); in full-duplex\footnote{In this context, full duplex typically refers to the user’s ability to interrupt the agent’s speech, whereas, in the strict sense, full duplex entails that the agent can likewise interrupt the user.} interaction, both parties can speak concurrently, allowing the user to interrupt or barge in while the agent is speaking.

Currently, voice interaction systems are implemented using three main paradigms. (1) Cascaded pipeline: an ASR (Automatic Speech Recognition) module transcribes the user’s speech into text, an LLM (Large Language Model) interprets the transcribed text and generates a textual reply, and a TTS (Text-to-Speech) module synthesizes the reply into speech. The advantage of this approach is modularity, each component optimized or replaced independently, allowing the use of diverse ASR~\cite{xu2025fireredasr, radford2023robust}, LLM~\cite{touvron2023llama, anil2023palm, chowdhery2023palm, team2024qwen2, yang2025qwen3}, and TTS~\cite{guo2024fireredtts, guo2025fireredtts, du2024cosyvoice, du2025cosyvoice, du2026cosyvoice, xie2025fireredtts} models. Its limitation is that the LLM conditions only on text and thus cannot exploit paralinguistic cues present in speech, such as emotion, speaker identity, and acoustic events. (2) Semi-cascaded pipeline: an AudioLLM~\cite{chu2023qwen, chu2024qwen2, tang2023salmonn, wu2023decoder, midashenglm7b} directly ingests speech and produces a textual response, which is then synthesized by TTS. Compared with the cascaded approach, this paradigm can leverage emotional and background acoustic information from the input, leading to more contextually appropriate textual responses. However, the downstream TTS typically does not condition on the user’s prosody and emotion, which are crucial for producing natural and stylistically consistent reply speech. (3) End-to-end speech-to-speech models~\cite{zhang2023speechgpt, du2023lauragpt, xie2024mini, fang2024llama, ding2025kimi, huang2025step, wu2025step, xu2025qwen2}: a single model directly maps user speech to response speech, maintaining awareness of emotion and background acoustics during both understanding and generation, while eliminating the intermediate text step. Nevertheless, this paradigm often faces two challenges: catastrophic forgetting in the LLM backbone and a length mismatch between text and speech token sequences, which complicates alignment as well as training and inference.

The aforementioned three voice interaction paradigms are inherently half-duplex. However, half-duplex interaction typically incurs higher latency and leads to a less convenient user experience. Full-duplex voice interaction can be realized in two ways: (i) handling turn-taking explicitly or implicitly within an end-to-end model, or (ii) relying on pluggable control modules to support barge-in, thereby promptly returning the conversational floor to the user. For the first approach, dGSLM~\cite{nguyen2023generative} is the first end-to-end full-duplex voice interaction model; it uses a single model to separately represent the user input stream and the model output stream, and employs cross-attention to enable mutual awareness between the two streams. Moshi~\cite{defossez2024moshi} sums the user input stream and the model output stream at the current time step to predict the next-step model output, and further introduces an Inner Monologue (a textual stream) to improve response quality. Freeze-Omni~\cite{wang2024freeze} and MinMo~\cite{chen2025minmo}, while not explicitly modeling dual streams, predict in real time, based on hidden states, whether to pause speech rendering or continue generation within an end-to-end speech-to-speech model. It is worth noting that end-to-end speech-to-speech models are already highly complex; adding duplex control typically requires additional architectural modifications and multi-stage, intricate training procedures, and any functional updates often necessitate retraining the entire model. Moreover, a salient drawback of end-to-end full-duplex systems is their limited controllability.

The second approach relies on pluggable control modules to orchestrate turn-taking. TurnGPT~\cite{ekstedt2020turngpt} injects a speaker embedding at each position and predicts the speaker ID to determine turn transitions. MThread~\cite{wang2024freeze}  uses LLM-predicted state tokens (Speak, Listen) to govern turn-taking. VITA~\cite{fu2024vita} deploys two models in paralle, one for response generation and another for continuous listening, and predicts a state token to manage turns. FlexDuo~\cite{liao2025flexduo} predicts turn transitions from past context and incoming speech chunks. With such control modules, any half-duplex system can be upgraded to full duplex while preserving modularity, enabling independent optimization and replacement of components. This underscores the need for a pluggable voice interaction framework capable of integrating heterogeneous models.

Several prior voice interaction frameworks, such as LiveKit~\footnote{\url{https://livekit.io}}, Ten~\footnote{\url{https://theten.ai}}, and Pipecat~\footnote{\url{https://www.pipecat.ai}}, adopt a modular, pluggable architecture that composes interchangeable components to construct end-to-end voice interaction pipelines, which is an effective design paradigm. However, some of these frameworks include components that are not fully open-sourced (e.g., denoising modules for suppressing input speech noise), which hinders holistic improvement and optimization of the entire interaction chain. Moreover, because these frameworks do not ship models themselves and instead rely on external providers, they do not expose a complete, end-to-end voice interaction pipeline along with optimization strategies.

\begin{figure}[h] \centering \includegraphics[width=1.0\textwidth]{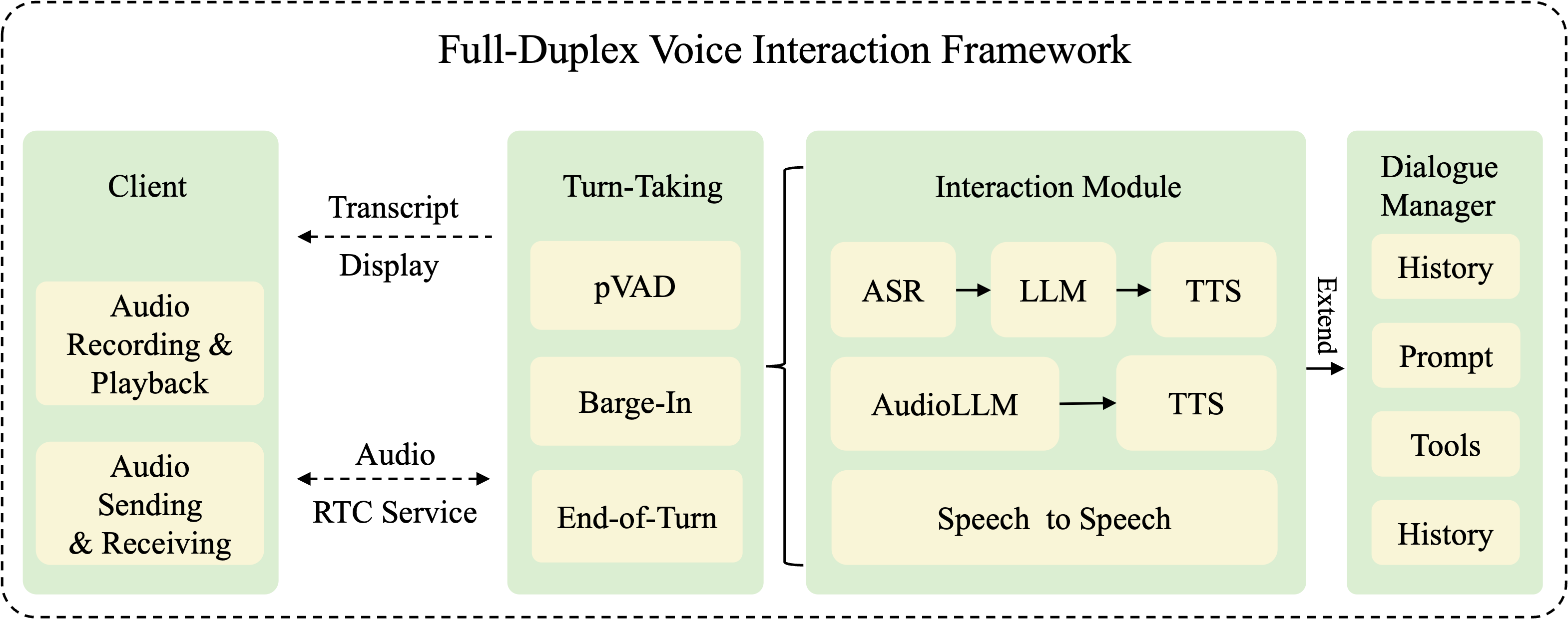} 
\caption{FireRedChat System Modules.} \label{fig:fireredcaht arc} \end{figure}

In this report, we present a complete, deployable full-duplex voice interaction system that encompasses all modules in the interaction pipeline (as illustrated in Figure~\ref{fig:fireredcaht arc}). The system comprises a dedicated turn-taking controller, an interaction module, and a dialogue manager. The turn-taking controller delivers human-like full-duplex interactions: the streaming personalized voice activity detection (pVAD) suppresses background noise and competing speakers to reduce false barge-ins, more precisely detects intentional interruptions initiated by the primary speaker, and produces accurate timestamps for the primary speaker’s speech segments. In addition, an End-of-Turn (EoT) component improves the system’s ability to detect when the user has semantically completed a turn.
With the turn-taking control module in place, any half-duplex interaction mode, cascaded, semi-cascaded, or speech-to-speech, can be integrated and upgraded to full duplex. Leveraging our in-house models, we provide both cascaded and semi-cascaded implementations. Cascaded architectures benefit from mature deployment practices and well-established pipeline optimization techniques. Semi-cascaded architectures, by contrast, enable both understanding and generation to be conditioned on the user’s emotional and paralinguistic cues, thereby supporting richer audio understanding and producing more contextually coherent responses. By replacing separate ASR and LLM stages, the semi-cascaded approach further reduces end-to-end latency, mitigates error propagation, and improves robustness.
To extend system capabilities, the dialogue manager supports tool invocation and context management. In addition, we introduce metrics for barge-in handling, end-of-turn detection, and latency to evaluate a system holistically along three dimensions: interaction naturalness, control accuracy, and efficiency.
Experimental results demonstrate that our system is more resilient to false barge-ins caused by non-target speakers and noise, achieves more accurate semantic end-of-turn detection, and narrows the interaction-latency gap with industrial-grade systems, thereby delivering a more robust, natural, and immediate full-duplex interaction experience.

\section{FireRedChat}

In our modular, plug-and-play full-duplex voice interaction system, we provide a dedicated turn-taking controller, an interaction module, and a dialogue manager. In the turn-taking controller, we employ a streaming personalized voice activity detection module (pVAD) to mitigate false interruptions caused by background noise and non-target speakers, to handle barge-in more precisely, and to more accurately localize the temporal boundaries of the primary speaker’s utterances. In addition, an end-of-turn (EoT) component performs more accurate semantic end-of-turn detection, effectively preventing premature truncation of user input and thereby improving the interaction experience.
Within the interaction module, cascaded, semi-cascaded, and speech-to-speech half-duplex pipelines can be integrated into our system to enable full-duplex operation. Using our in-house models, we currently support both cascaded and semi-cascaded configurations. Specifically, the cascaded pipeline comprises FireRedASR~\cite{xu2025fireredasr}, Qwen2.5, and FireRedTTS-1s~\cite{guo2025fireredtts}, whereas the semi-cascaded pipeline uses AudioLLM and FireRedTTS-2~\cite{xie2025fireredtts}. The cascaded approach benefits from technological maturity, while the semi-cascaded approach, during both understanding and generation, can perceive emotional cues in the user’s speech and other acoustic conditions, enabling more appropriate responses and delivering a lifelike interaction experience.
To further extend the system’s comprehension capabilities, we integrate a dialogue manager via Dify~\footnote{\url{https://dify.ai}}, which supports tool invocation and context management.

\subsection{Workflow}

\begin{figure}[h] \centering \includegraphics[width=0.7\textwidth]{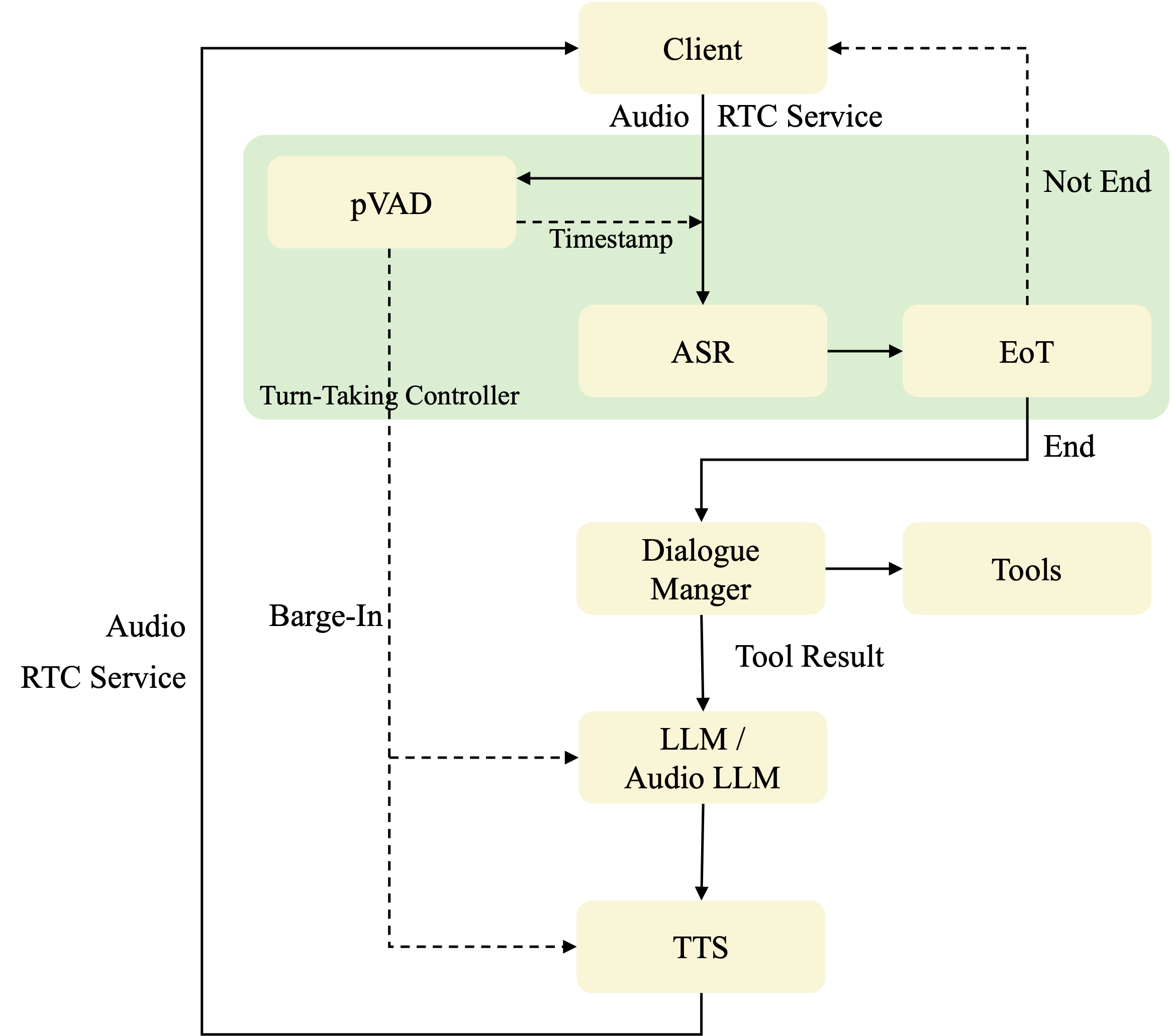} 
\caption{FireRedChat Voice Interaction Flow.} \label{fig:fireredcaht flow} \end{figure}

As illustrated in Figure~\ref{fig:fireredcaht flow}, the user’s speech is transmitted via RTC to the agent. The incoming audio is first analyzed by our streaming pVAD module, which suppresses background noise and non-target speakers, yields precise timestamps for the primary speaker’s segments, and improves detection of whether the user is speaking. This, as a result, increases the success rate of user-initiated barge-in while reducing spurious interruptions caused by ambient noise or other speakers. Whenever the primary speaker is detected, the system immediately suspends TTS playback, thereby enabling a full-duplex interaction experience.
We employ pVAD exclusively for primary-speaker time-stamping and robust barge-in control; we do not feed pVAD-denoised segments to downstream components. Instead, we use the pVAD timestamps to extract the corresponding segments from the user’s original, non-denoised audio. This design choice serves two purposes: (i) it preserves ambient and other paralinguistic cues for downstream perception, and (ii) LLM-based ASR models, exemplified by our FireRedASR, are trained on large-scale data and are intrinsically robust to noise and speaker overlap, so recognizing directly from the original signal has negligible adverse impact in practice.
The non-denoised speech segments are forwarded to the ASR module to obtain transcripts. The EoT component then performs semantic stop detection: if the utterance has not semantically ended, partial transcripts are buffered while the system continues to accept new audio input until an end-of-turn is reached. The complete transcript is passed to the dialogue manager, which decides whether to invoke tools (e.g., WebSearch). When propagating tool outputs to downstream modules, the results are typically prefixed with phrases such as "You may refer to the following content:". Subsequent response generation can proceed in cascaded, semi-cascaded, or speech-to-speech configurations. In the cascaded setup, a text LLM consumes both the user transcript and any tool results to produce a response; in the semi-cascaded setup, an AudioLLM consumes the user’s speech together with the tool results to generate a response. Finally, the response is synthesized by the TTS module for playback.

\subsection{Turn-taking Controller}

\subsubsection{Streaming Personalized VAD}

In full-duplex voice interaction, systems are often spuriously interrupted by non-primary speakers or fail to promptly honor barge-in by the primary speaker, which degrades user experience. To address this, we introduce a streaming personalized voice activity detection module (pVAD) that suppresses background noise and non-target speakers, thereby reducing false interruptions, improving the success rate of primary-speaker barge-in, and yielding more accurate localization of the target speaker’s speech segments.

Specifically, building upon~\cite{vafeiadis2019two}, we incorporate target-speaker conditioning and redesign the model for streaming inference. For each incoming audio chunk, we compute mel-spectrogram features and process them with causal convolution layers to ensure strict temporal causality. A target-speaker embedding extracted by an ECAPA-TDNN~\cite{desplanques2020ecapa} encoder is concatenated with the post-convolution features along the channel dimension, after which a GRU~\cite{dey2017gate} module models temporal dependencies. A final classifier outputs the speaking probability at each time step; each time step corresponds to a $10$ms audio chunk.
We curated a total of $2000$ hours of clean Mandarin and English speech. During training, we construct mixtures by (i) combining a $5$s target-speaker segment with a $5$s interfering-speaker segment at a signal-to-noise ratio (SNR) randomly sampled from $0$ to $30$ dB, and (ii) combining a $5$s target-speaker segment with a $5$s noise segment at an SNR sampled from the same range. Interfering speakers and noise~\cite{snyder2015musan} are injected with equal probability ($50\%$). The resulting streaming pVAD operates without waiting for the complete utterance, enabling earlier responses to user barge-ins while remaining robust to spurious activations induced by interfering speakers and background noise.


\subsubsection{End-of-Turn Detection}

A key distinction between speech-based and text-based interaction is that text entry offers an explicit “send” action, whereas speech requires the system to infer whether the user has reached a semantic end-of-turn. Conventional VAD can indicate that a contiguous acoustic segment has ended, but this does not necessarily imply that the user has finished their input. To bridge this gap, our end-of-turn (EoT) module assesses all accumulated ASR transcripts and predicts whether a semantic stop has been reached; only upon such a decision does the pipeline advance to the next stage.

Concretely, we fine-tune a classifier on top of a pre-trained BERT~\footnote{\url{https://huggingface.co/google-bert/bert-base-multilingual-cased}} model for semantic stop detection. We construct a corpus of approximately 830,000 text instances by sampling partial spans from complete utterances to simulate unfinished states, paired with full utterances representing completed states. The EoT module thus learns to discriminate “continue” versus “stop” given the concatenated ASR hypotheses observed so far. By making end-of-turn decisions at the semantic level, EoT renders voice interaction more conversational, akin to human dialogue, determining when it is appropriate to respond based on the interlocutor’s progress within an utterance.

\subsection{Interaction Module}

\subsubsection{Cascade Implementation}

In the cascaded configuration, our ASR component is FireRedASR, which achieves strong performance on Mandarin and English speech recognition across a range of public and proprietary datasets. We use Qwen2.5 as the LLM. For TTS, we adopt FireRedTTS-1s, which supports streaming decoding to reduce end-to-end latency and provides distinctive speaker timbres, yielding natural-sounding synthesis. 
A key advantage of the cascaded approach is that industry practices for its deployment and pipeline optimization are already well established.
Moreover, to further lower overall system latency, we implemented engineering optimizations that accelerate both FireRedASR and FireRedTTS-1s.

\subsubsection{Semi-cascade implementation}

In the semi-cascaded configuration, we deploy an AudioLLM together with FireRedTTS-2. The AudioLLM consumes the user’s speech (optionally augmented with tool outputs) and produces a textual reply. FireRedTTS-2 then synthesizes the response speech while conditioning on the user’s input audio as contextual guidance. Compared with the cascaded pipeline, the semi-cascaded approach offers several advantages:

\begin{enumerate}
    \item Paralinguistic awareness: Both AudioLLM and FireRedTTS-2 can perceive non-lexical attributes in the user’s speech (e.g., emotion). By jointly considering semantics and acoustics, they adapt response style accordingly and yield more natural interactions.
    \item Reduced error propagation: AudioLLM replaces the ASR+LLM stages and jointly models acoustic and linguistic information in the input speech, thereby mitigating ASR error cascades and improving overall robustness.
    \item Audio understanding capability: Unlike cascaded systems, which discard acoustic evidence after ASR, AudioLLM can perceive audio events (e.g., human voices, music, ambient sounds) and support tasks requiring audio understanding, enabling a more lifelike user experience.
    \item Simpler pipeline with potential latency benefits: By subsuming ASR and LLM into a single model, AudioLLM simplifies the processing chain, potentially reducing end-to-end latency and engineering complexity. Relative to end-to-end speech-to-speech systems, decoupling understanding (AudioLLM) from generation (TTS) allows independent optimization; learning both within a single model often complicates training.
    \item Consistent paralinguistic conditioning in synthesis: When generating the spoken response, FireRedTTS-2 conditions on the user’s input speech, capturing emotional tone and background characteristics so that both “listening/thinking” (AudioLLM) and “speaking” (FireRedTTS-2) are aligned to the same semantic and acoustic cues, producing more coherent and consistent replies.
\end{enumerate}

AudioLLM is an in-house model trained from scratch following the Qwen2-Audio~\cite{chu2024qwen2} architecture. To endow it with perception of both paralinguistic and semantic information, we collect large-scale corpora spanning ASR, speech emotion recognition, acoustic scene classification, vocal sound classification, and audio captioning. After establishing broad audio understanding capabilities, we construct 250k topic-specified, colloquial dialogue instances synthesized via TTS, which teach the model to produce contextually appropriate, conversational responses grounded in the user's speech.

FireRedTTS-2 structures training data as interleaved user–system turns, which naturally conforms to dialog dynamics. It is first trained on 1.1M hours of single-sentence speech, then further trained on 300k hours of dialog data, and finally fine-tuned on a small set of signature timbres. This design makes FireRedTTS-2 particularly strong in dialog-style generation and enables it to re-condition on the user’s speech during synthesis to produce emotionally consistent, context-aware responses.

\section{Evaluation}

To evaluate voice interaction systems from the perspective of overall interaction experience, we introduce three metrics: barge-in, end-of-turn detection, and latency, which respectively quantify interaction naturalness, interaction control accuracy, and interaction efficiency.
To compare framework-specific modules (e.g., VAD, EoT) and to enable the fairest possible comparison of latency attributable to ASR and TTS, we standardized the LLM component to Qwen2.5 across LiveKit, Ten, and our system. The ASR, TTS, and control modules were configured using each framework’s best-performing settings. Detailed configurations are provided in Table~\ref{tab:config}. We accessed DouBao via its official mobile application.
Experimental results indicate that our system exhibits improved robustness to false barge-ins induced by other speakers and background noise, delivers more accurate semantic end-of-turn detection of user input, and narrows the interaction-latency gap relative to industrial-grade applications, thereby enabling a more robust, natural, and real-time full-duplex interaction experience.

\begin{table}[h]
\centering
\caption{Configurations between different systems.}
\label{tab:config}
\begin{tabular}{@{}cc@{}}
\toprule
System                   & Component \\ \midrule
\multirow{5}{*}{LiveKit} & SileroVAD~\tablefootnote{\url{https://pypi.org/project/livekit-plugins-silero}}  \\
                         & LiveKit EoT~\tablefootnote{\url{https://pypi.org/project/livekit-plugins-turn-detector}}        \\
                         & Tencent ASR~\tablefootnote{\url{https://cloud.tencent.com/document/product/1093/35646}}        \\
                         & Qwen2.5~\tablefootnote{\url{https://huggingface.co/Qwen/Qwen2.5-7B-Instruct}}        \\
                         & DouBao TTS~\tablefootnote{\url{https://www.volcengine.com/docs/6561/1257584}}        \\ \midrule
\multirow{5}{*}{Ten}     & Ten VAD~\tablefootnote{\url{https://github.com/TEN-framework/ten-vad}}    \\
            & Ten EoT~\tablefootnote{\url{https://github.com/TEN-framework/ten-turn-detection}} \\
                         & Azure ASR~\tablefootnote{\url{https://learn.microsoft.com/en-us/azure/ai-services/speech-service/speech-to-text}} \\
                         & Qwen2.5       \\
                         & DouBao TTS        \\ \midrule
\multirow{5}{*}{FireRedChat}
                         & FireRedChat pVAD       \\
                         & FireRedChat EoT \\
                         & FireRedASR~\tablefootnote{\url{https://github.com/FireRedTeam/FireRedASR}} \\
                         & Qwen2.5       \\
                         & FireRedTTS-1s~\tablefootnote{\url{https://github.com/FireRedTeam/FireRedTTS}}     \\ \bottomrule
\end{tabular}
\end{table}

\subsection{Barge-In}

Barge-in denotes the requirement that a system immediately cease TTS playback once the user begins speaking. To achieve a lifelike interaction experience, the system must reliably distinguish the primary speaker from non-primary speakers and background sounds, correctly detect the onset of user speech, and promptly halt ongoing playback.

We evaluate barge-in using two metrics: (i) For the barge-in success rate, we compute accuracy at specific offsets relative to the onset of user speech (e.g., $0$\,ms, $50$\,ms, $100$\,ms). We then report the minimum latency required to reach 90\% barge-in accuracy (denoted $T_{90}$), which quantifies how quickly the system responds to user interruptions. (ii) The false barge-in rate measures erroneous interruptions triggered when the primary speaker is silent; in such cases, any interruption induced by other sounds (background noise or non-primary speakers) is counted as an error. This metric reflects the system’s robustness to non-speech and non-primary-speaker interference.

To assess these metrics, we constructed $1{,}000$ utterances each in Chinese and English. To simulate noise and competing speakers during user speech, background noise was injected with $50\%$ probability prior to the onset of user speech, ensuring overlap around the speech-onset region. Noise samples were drawn from~\cite{dubey2024icassp} and comprise acoustic scenes such as vehicular and office environments. To disentangle the effects of background noise from competing talkers, we additionally inserted a secondary speaker within a $[-1, 1]$ s window around the end of the primary speaker.
The SNR between the primary speaker and the background noise was fixed at $5$ dB, whereas the SNR between the primary and secondary speakers was fixed at $20$ dB.

The results are reported in Table~\ref{tab:barge-in}.
Enabled by our streaming pVAD, the system substantially mitigates false barge-ins caused by interfering speakers and background noise, while still responding promptly to target-speaker barge-ins, thereby improving the overall interaction experience. Although Ten achieves a very low $T_{90}$, this also indicates that the system is highly sensitive to any acoustic event, which leads to an elevated false barge-in rate. Furthermore, because neither LiveKit nor Ten incorporate target-speaker conditioning, their higher false barge-in rates are expected. Compared with LiveKit, our method exhibits a roughly 30 ms higher $T_{90}$; however, this modest additional wait helps suppress false barge-ins and thus represents a worthwhile trade-off.

\begin{table}[h]
\centering
\caption{Barge-In evaluation results.}
\label{tab:barge-in}
\begin{tabular}{@{}ccc@{}}
\toprule
System             & $T_{90}$ (ms) & False barge-in rate (\%) \\ \midrule
LiveKit &        140              &          33.4         \\
Ten &         \textbf{90}              &         78.1            \\
FireRedChat         &         170              &            \textbf{10.2}         \\ \bottomrule
\end{tabular}
\end{table}

\subsection{End-of-Turn Detection}

End-of-turn detection refers to correctly identifying when the user’s spoken input has semantically concluded. Crucially, this boundary is semantic rather than purely acoustic: it does not coincide with the tail of a VAD segment, but with the completion of the user’s intended message. This differentiates voice interaction from text-based interfaces, where users explicitly signal completion via a “send” action. In voice interaction, the system must decide when the user has finished speaking, making accurate EoT detection central to turn-taking and interaction control.

For evaluation, we use the official test set from the TEN\_Turn\_Detection~\cite{TEN_Turn_Detection}. This set contains texts that include a semantic stopping point as well as texts that do not. We formulate EoT as a binary decision task (stop vs. continue) and report the decision accuracy over the test set. The result is shown in Table~\ref{tab:end-of-turn}.
For both Chinese and English, our model delivers superior EoT detection performance compared with LiveKit at a comparable parameter scale, and achieves parity with Ten's 7B-parameter model.
This design allows our interaction pipeline to maintain sufficiently accurate interaction control while imposing negligible overhead on end-to-end latency, thereby yielding a smoother and more natural user experience.

\begin{table}[h]
\centering
\caption{End-of-turn Detection evaluation results.}
\label{tab:end-of-turn}
\begin{tabular}{@{}cccccc@{}}
\toprule
System                  & \#Params              & Language & Finished (\%) & Unfinished(\%) & Average(\%) \\ \midrule
\multirow{2}{*}{LiveKit} & \multirow{2}{*}{373M} & Chinese  & 92.6          & 45.3           & 70.8      \\
                         &                       & English  & 76.3          & 98.4           & 86.2      \\ \midrule
\multirow{2}{*}{Ten}     & \multirow{2}{*}{7B}   & Chinese  & 98.5 & 92.7           & 95.8     \\
                         &                       & English  & 91.1          & 98.4           & 94.4      \\ \midrule
\multirow{2}{*}{FireRedChat}    & \multirow{2}{*}{170M} & Chinese  & 96.3          & 95.7  & \textbf{96.0}     \\
                         &                       & English  & 96.2          & 93.2           & \textbf{94.9}    \\ \bottomrule
\end{tabular}
\end{table}

\subsection{Latency}

We define latency as the end-to-first-response time required to complete an interaction turn at the client side. Concretely, latency is the elapsed time from the end of the user’s input audio to the moment the client receives the first audio chunk produced by the system. This definition is system-agnostic: it applies equally to cascaded and end-to-end architectures and serves as a direct measure of overall processing efficiency. A responsive system should comprehend the user’s input and provide timely feedback as soon as possible.

Specifically, to ensure consistency of the input audio, we selected 25 representative synthetic utterances from the constructed dialogue dataset, covering variations in emotion and pauses, with moderate durations. We then screen-recorded the interaction sessions and computed the latency from the end of the audio input to the first audible response from the interactive system. Notably, the position and volume of the phone used to play the audio were kept constant across all experiments. These audio inputs do not trigger tool invocation.
Given that audio input lengths vary, we report the 50th-percentile ($P_{50}$) latency as a proxy for the typical user experience and the 95th-percentile ($P_{95}$) latency to characterize the long tail. 
The result is shown in Table~\ref{tab:latency}.
Compared with the open-source frameworks LiveKit and Ten, our system achieves $P_{50}$ latencies lower by 1.2 s and 1.0 s, respectively, and $P_{95}$ latencies lower by 1.6 s and 0.8 s. By contrast, relative to the industrial system DouBao, our $P_{50}$ and $P_{95}$ latencies are higher by only 0.3 s and 0.6 s, respectively.
We believe that the observed latency disparity can be further narrowed or even eliminated by enabling streaming output in ASR, streaming input in TTS, and further reducing the VAD audio chunk size.

\begin{table}[h]
\centering
\caption{Latency of different systems.}
\label{tab:latency}
\begin{tabular}{@{}ccc@{}}
\toprule
System  & $P_{50}$ (s) & ${P_{95}}$ (s) \\ \midrule
LiveKit &   3.598    &   4.649    \\
Ten     &    3.375    &   3.802   \\
DouBao    &   \textbf{2.075}    &   \textbf{2.407}  \\ 
FireRedChat    &    2.341     &  3.015   \\\bottomrule
\end{tabular}
\end{table}

\section{Conclusions}

In this report, we present a complete, deployable solution for full-duplex voice interaction. The system comprises a dedicated turn-taking controller, an interaction module, and a dialogue manager. The turn-taking controller enables lifelike, continuous conversation via a streaming personalized VAD (pVAD) and an end-of-turn (EoT) detector: pVAD reduces false interruptions from background noise and non-target speakers, enables precise detection of barge-ins by the primary speaker, and yields accurate timestamps for primary-speaker segments, while the EoT module identifies the semantic completion of user turns.
Built on this controller, any half-duplex pipeline, cascaded, semi-cascaded, or speech-to-speech, can be upgraded to full duplex. Leveraging our internal models, we provide both cascaded and semi-cascaded implementations. The cascaded design benefits from mature deployment and optimization practices; the semi-cascaded design perceives emotional and paralinguistic cues in both understanding and generation, broadens audio understanding, improves contextual coherence, and by reducing reliance on separate ASR and LLM stages, lowers latency, mitigates error propagation, and enhances robustness.
To extend functionality, the dialogue manager supports tool invocation and context management. Finally, we introduce interruption, end-of-turn detection, and latency metrics that assess interaction naturalness, control accuracy, and efficiency.
Experimental results indicate that our system better resists spurious interruptions induced by other speakers and background noise, achieves more precise semantic end-of-turn detection, and further narrows the interaction-latency gap relative to industrial-grade applications, thereby delivering a more robust, natural, and real-time full-duplex interaction experience.

\bibliographystyle{unsrt}
\bibliography{refs}

\end{document}